\documentstyle[12pt]{article}

\begin{document}
\setcounter{page}{0}

\newcommand{\rf}[1]{(\cite{#1})}
\newcommand{\R}{{\rm I\!R}}
\newcommand{\Z}{{\sf Z\!\!Z}}
\newcommand{\goto}{\rightarrow}

\newcommand{\al}{\alpha}
\renewcommand{\b}{\beta}
\renewcommand{\c}{\chi}
\renewcommand{\d}{\delta}
\newcommand{\D}{\Delta}
\newcommand{\ve}{\varepsilon}
\newcommand{\f}{\phi}
\newcommand{\F}{\Phi}
\newcommand{\vf}{\varphi}
\newcommand{\g}{\gamma}
\newcommand{\G}{\Gamma}
\newcommand{\k}{\kappa}
\renewcommand{\l}{\lambda}
\renewcommand{\L}{\Lambda}
\newcommand{\m}{\mu}
\newcommand{\n}{\nu}
\newcommand{\r}{\rho}
\newcommand{\vr}{\varrho}
\renewcommand{\o}{\omega}
\renewcommand{\O}{\Omega}
\newcommand{\p}{\psi}
\renewcommand{\P}{\Psi}
\newcommand{\s}{\sigma}
\renewcommand{\S}{\Sigma}
\newcommand{\th}{\theta}
\newcommand{\vt}{\vartheta}
\renewcommand{\t}{\tau}
\newcommand{\vp}{\varphi}
\newcommand{\x}{\xi}
\newcommand{\z}{\zeta}
\newcommand{\ta}{\triangle}
\newcommand{\w}{\wedge}
\newcommand{\e}{\eta}
\newcommand{\Th}{\Theta}
\newcommand{\td}{\tilde}
\newcommand{\ep}{\epsilon}
\newcommand{\na}{\nabla}
\newcommand{\be}{\begin{equation}}
\newcommand{\ee}{\end{equation}}
\newcommand{\bE}{\begin{eqnarray}}
\newcommand{\eE}{\end{eqnarray}}
\newcommand{\nn}{\nonumber}
\renewcommand{\thefootnote}{\fnsymbol{footnote}}

\newcommand{\siml}{\raisebox{-.6ex}{$\stackrel{<}{\displaystyle{\sim}}$}}
\newcommand{\simg}{\raisebox{-.6ex}{$\stackrel{>}{\displaystyle{\sim}}$}}
\newcommand{\ind}{\scriptscriptstyle}

\newcommand{\PR}{Phys. Rev. }
\newcommand{\PRL}{Phys. Rev. Lett. }
\newcommand{\NP}{Nucl. Phys. }

\newpage
\setcounter{page}{0}
\begin{titlepage}
%February 1994
\begin{flushright}
\hfill{KAIST-CHEP-94-07}\\
\vskip 2mm
\hfill{YUMS-94-05}
\end{flushright}
\vspace{0.7cm}
\begin{center}
{\Large\bf Polyakov's Spin Factor for a Classical Spinning}
\vskip 3mm
{\Large\bf Particle via BRST Invariant Path Integral}

\vskip 1.5cm
{\bf Jin-Ho Cho \footnote{jhcho@chiak.kaist.ac.kr}}
\vskip 0.3cm
{\sl Department of Physics\\
Korea Advanced Institute of Science and Technology\\
373-1 Yusung-ku, Taejon, 305-701, Korea}
\vskip 0.5cm
{\bf Seungjoon Hyun \footnote{hyun@phya.yonsei.ac.kr} and
Hyuk-jae Lee \footnote{lhjae@bubble.yonsei.ac.kr}}
\vskip 0.3cm
{\sl Institute for Mathematical Sciences\\
Yonsei University, Seoul, 120-749, Korea}
\vskip 0.5cm

\end{center}

\setcounter{footnote}{0}

\begin{abstract}
%\noindent
For the `classical' formulation of a massive spinning particle,
the propagator is obtained along with the spin factor. We treat the
system with two kinds of constraints that were recently shown
to be concerned with the reparametrization invariance and
`quasi-supersymmetry'. In the path integral, the
BRST invariant Lagrangian is used and the same spin factor is obtained as in
the pseudo-classical formulation.
\end{abstract}

\end{titlepage}

\newpage
\renewcommand{\thefootnote}{\arabic{footnote}}
%\baselineskip=20pt
%\section{Introduction}

There are two standard ways of describing the
spin degrees of freedom for the relativistic particle; the `classical' way
describes them in terms of
the Lorentz group elements \cite{han} and the `pseudo-classical' one does
that in terms of the Grassmann quantities
\cite{rbe}. The main difference between them is the
symmetries. The classical system for the relativistic particle has the
reparametrization symmetry while the pseudo-classical one has extra
supersymmetry.

In Ref. \cite{cho}, it is shown that this difference is due to the
constraint structure of the systems; the constraint $p_\m\cdot S^{\m\n}=0$
for the classical system \cite{rba} can be relaxed without changing the
physical properties of the system. Furthermore that constraint, which can be
considered as the supersymmetry analogue of the classical system, gives a
closed algebra together with the constraint $p^2+m^2=0.$

In this letter, we are to confirm again that the model gives the right
description of the relativistic spinning particle through
the BRST invariant construction of the propagator along with the spin
factor.

The model is given by the following Lagrangian
\bE
{\cal L}&=& p^\m \dot x_\n-\frac{\l}{2}(t_1\dot t_2-t_2 \dot
t_1)-\frac{\l}{2}(\L_{\m 1}\dot\L^\m{}_2
-\L_{\m 2}\dot\L^\m{}_1)\nn \\
& &-M_1(\frac{p_\m}{m}\L^\m{}_2
+t_2)+M_2(\frac{p_\m}{m}\L^\m{}_1
+t_1)-N(p^2+m^2)\nn \\
&\equiv& p^\m\dot x_\m
-\l t_1 \dot t_2 -\l u \cdot\dot v -M_1\F_1
-M_2\F_2 -N\F_N,
\label{aaa}
\eE
where $u^2=1, v^2=1$ and $u\cdot v=0$ \footnote{we set $\l^{\m\n}$ of
\cite{cho} equal to $\l \d^\m_1\d^\n_2$ for our convenience.}.
{}From this first order Lagrangian the Poisson brackets are
easily shown to be

\bE
\{ x^\m,p^\n \}&=&\eta^{\m\n},\;\;\; \{ t_1,t_2\}=\frac{1}{\l},\nn \\
\{ u^\m,v^\n \}&=&\frac{1}{\l}\eta^{\m\n},
\label{aab}
\eE
and the constraint algebra is given by
\bE
\{ \F_N,\F_N \}&=&\{ \F_N,\F_i\}=0\nn \\
\{ \F_i,\F_j \}&=&-\frac{\ep_{ij}}{\l m^2}\F_N,\;\;\; i,j=1,2.
\label{con}
\eE
They form a closed algebra similar to the supersymmetry of the
pseudo-classical formulation. Here, we do not treat the group relation
$u^2=1,v^2=1$ and $u\cdot v=0$ as the `dynamical' constraints
\footnote{in ref. \cite{cho}, they are thus dealt with and result in similar
quasi-supersymmetry although in a somewhat complicated form.}.
Being absorbed into the measure of the path integral they do the role of
reducing the integration range.

The Lagrangian can be written in the second order style
\be
{\cal L}=-m\sqrt{-(\dot x - \frac{\l}{m}(v\dot t_1-u\dot t_2))^2}
+\l t_1\dot{t}_2 -\l u\cdot\dot v,
\label{seo}
\ee
in which $N^2=-\frac{1}{4m^2}(\dot x - \frac{\l}{m}(v\dot t_1-u\dot
t_2))^2$ (this comes from the equation of motion for $p$) is used.
We choose $N$ to be positive for later use.

%\section{BRST formulation}
In order to deal with the gauge invariant system given above,
we follow the standard BRST formulation \cite{brst}.
For the constraints above (\ref{con}), we straightforwardly
construct the BRST operator \cite{bvf}
\be
Q=c_N \F_N + c_i\F_i +\frac{1}{\l m^2}c_1 c_2 \c_N +\pi_N\bar\c_N
+\pi_1\bar\c_1+\pi_2\bar\c_2,\;\;i=1,2
\label{q}
\ee
and the gauge fermion
\be
\p =\c_N N+\c_1 M_1 +\c_2 M_2,
\label{gf}
\ee
which is corresponding to the usual gauge fixing $\dot N=\dot M_1
=\dot M_2=0$.
Here $(c_a,\bar \c_a)$ and $(\c_a, \bar c_a)$ are ghosts and anti-ghosts,
respectively. We use the following Poisson brackets for ghost variables:
\bE
\{ N,\pi_N\}&=&\{ M_i,\pi_i\} =1,\;\;i=1,2\nn \\
\{ c_a,\c_a\}&=&1,\;\; \{ \bar c_a,\bar\c_a\} =1,\;\; a=1,2,N.
\label{pbg}
\eE
The BRST invariant Hamiltonian is
\bE
H_Q &=& \{ \p,Q \} \nn\\
    &=& N\F_N+M_1\F_1+M_2\F_2+\frac{1}{\l m^2}c_2 M_1\c_N \nn\\
    & &-\frac{1}{\l m^2}c_1 M_2\c_N+\c_N\bar\c_N+\c_1\bar\c_1+\c_2\bar\c_2.
\label{H}
\eE

%\section{BRST invariant path integral}
The effective BRST invariant Lagrangian is obtained through the Legendre
transformation as follows
\bE
{\cal L}_{eff}&=& p\cdot\dot x-\l t_1 \dot t_2 -\l u\dot v +\dot N\pi_N
+\dot M_1\pi_1+\dot M_2\pi_2 \nn\\
& &+\dot c_N\c_N+ \dot c_1\c_1+ \dot c_2\c_2 - \dot {\bar
c}_N\bar\c_N- \dot{\bar c}_1\bar\c_1 -\dot{\bar c}_2\bar\c_2-H_Q,
\label{bla}
\eE
where the canonical kinetic terms for the ghost variables are assumed. For
later convenience we
divide the whole Lagrangian into two parts; the matter part
\bE
{\cal L}_{matter}&=&p\cdot\dot x-\l t_1 \dot t_2 -\l u\dot v +\dot N\pi_N
+\dot M_1\pi_1+\dot M_2\pi_2 \nn\\
& &-N(p^2+m^2)-M_1(\frac{p}{m}\cdot v+t_2)-
M_2(\frac{p}{m}\cdot u+t_1),
\label{lma}
\eE
and the ghost part
\bE
{\cal L}_{ghost}&=&\dot c_N\c_N+ \dot c_1\c_1+ \dot c_2\c_2 - \dot{\bar
c}_N\bar\c_N- \dot{\bar c}_1\bar\c_1 -\dot{\bar c}_2\bar\c_2 \nn \\
& &-\frac{1}{\l m^2}c_2
M_1\c_N+\frac{1}{\l m^2}c_1 M_2\c_N-\c_N\bar\c_N-\c_1\bar\c_1-\c_2\bar\c_2.
\label{lgh}
\eE

Now we are to consider transition amplitude via Feynman path integral.
The matter part is written as
\bE
Z_{matter}&=&\int DpDxDt_1 Dt_2 \tilde Du\tilde DvDNDM_1 DM_2 D\pi_N
D\pi_1 D\pi_2 \d[u\cdot v]\nn \\
& &\cdot\exp\{ -i\int d\t [p\cdot\dot x-\l t_1 \dot t_2 -\l u\dot v
+\dot N\pi_N +\dot M_1\pi_1+\dot M_2\pi_2 \nn\\
& &-N(p^2+m^2)-M_1(\frac{p}{m}\cdot v+t_2)-M_2(\frac{p}{m}\cdot u+t_1)]\},
\label{zma}
\eE
where $\tilde Du\tilde Dv=DuDv\d[u^2-1]\d[v^2-1].$
The integration over $\pi_a,\;a=1,2,N$ gives the delta functionals
$\d[\dot N],\d[\dot M_1]$ and $\d[\dot M_2],$ which result in converting
those functional measures $DNDM_1 DM_2$ into the ordinary ones
$dNdM_1dM_2.$ The same is true for the integration over $x$ and the measure
$Dp$ turns into $dp.$ The ordinary integration over $N$, which is considered
to be positive, can be performed straightforwardly to
give the usual propagator for the relativistic particle with appropriate
regularization.
\be
\int dN e^{-i\int d\t N(p^2+m^2)} =\int dN
\exp\{ -iTN(p^2+m^2)\}=\frac{-i}{(p^2+m^2-i\ve)T,}
\label{prop}
\ee
where $T=\t_f-\t_i.$
Now the matter part may be rewritten as
\bE
Z_{matter}&=&\int dp dM_1 dM_2 Dt_1 Dt_2 \tilde Du \tilde Dv \d[u\cdot v]
\frac{i}{(p^2+m^2-i\ve)T}\nn \\
& &\cdot\exp\{ -i\int d\t [\l t_1 \dot t_2+\l u\cdot \dot v
+M_1\F_1+M_2\F_2]\}.
\label{mat}
\eE

The ghost part reads as
\bE
Z_{ghost}&=& \int DcD\c D\bar c D\bar \c  \nn\\
& &\cdot\exp\{ i\int d\t [\dot c_a\c_a - \dot{\bar c}_a\bar\c_a-\c_a\bar\c_a
-\frac{1}{\l m^2}c_2 M_1\c_N+\frac{1}{\l m^2}c_1 M_2\c_N]\} \nn \\
&=&\int DcD\c D\bar c D\bar\c\nn\\
& &\cdot\exp\{ i\int d\t [-(\c_a +\dot{\bar c}_a)(\bar \c_a +\dot c_a
-f^a_{bc}c_b M_c)+\dot{\bar c}_a(\dot c_a -f^a_{bc}c_b M_c)]\},\nn\\
\label{zgh}
\eE
where the structure constant $f^a_{bc}$ is nonvanishing only for the
following component
\be
f^N_{ij}=-\frac{\ep_{ij}}{\l m^2}.
\label{f}
\ee
This Gaussian integration can be performed easily by the field shifting
\bE
Z_{ghost}&=&\int DcD\c D\bar c D\bar\c \exp\{ i\int d\t (-\c_a \bar \c_a
+\dot{\bar c}_a\dot c_a) \} \nn\\
&=&const (iT)^3.
\label{gauss}
\eE
We note that this ghost part has only $T$ dependency.

Now we are in a position to finish the remaining integration for the matter
part.
\bE
Z_{matter}&=&\int dp dM_1 dM_2 Dt_1 Dt_2 \tilde Du \tilde Dv \d[u\cdot v]
\frac{-i}{(p^2+m^2-i\ve)T}\nn \\
& &\cdot\exp\{ -i\int d\t [\l t_1 \dot t_2+\l u\cdot \dot v
+M_1(\frac{p}{m}\cdot v+t_2)+M_2(\frac{p}{m}\cdot u+t_1)]\}. \nn\\
\label{euv}
\eE
The integration over $t_1$ gives delta functional
\be
\int Dt_1 e^{-i \int d\t t_1 (\l \dot t_2 +M_2)} =\d [\l\dot t_2 +M_2],
\ee
which converts the functional integral over $t_2$ into the ordinary one over
the constant mode $t_{2i}$ given as
\be
t_2(\t)=-\frac{M_2}{\l}\t +\frac{M_2}{\l}\t_i +t_{2i}.
\label{mode}
\ee
With the substitution of (\ref{mode}), the matter part is summarized as
\bE
Z_{matter}&=&\int dp dM_1 dM_2 dt_{2i} \tilde Du \tilde Dv \d[u\cdot v]
\frac{-i}{(p^2+m^2-i\ve)T}\nn \\
& &\cdot e^{ -i\int d\t [\l u\cdot \dot v +M_1(-\frac{M_2}{\l}\t
+\frac{M_2}{\l}\t_i +t_{2i})- M_1\frac{p}{m}\cdot v
+M_2\frac{p}{m}\cdot u]}. \nn\\
\label{zzz}
\eE

Finally the integrations over $t_{2i}$ and $M_1$ result in
\be
\int dt_{2i}\exp\{ -i M_1 T t_{2i} \} =\d (M_1 T)
\ee
and
\be
\int dM_1 \d(M_1 T)\exp\{ -i\int d\t M_1(-\frac{M_2}{\l}\t
+\frac{M_2}{\l}\t_i+t_{2i})\} =\frac{1}{T}.
\ee
Furthermore $\dot v$ may be expanded as $\dot v=C_{21}u+C_{11}v+w,$ where
$u\cdot w=v\cdot w=0,$ thus the matter part becomes
%\newpage
\bE
Z_{matter}&=&\int dp dM_2 \tilde Du \tilde Dv \d[u\cdot v]\nn\\
& &\cdot\frac{-i}{(p^2+m^2-i\ve)T^2}\exp\{ -i\int d\t
[\l u\cdot \dot v+M_2\frac{p}{m}\cdot u]\} \nn\\
&=&\int dp dM_2 \tilde Du \tilde Dv \d[u\cdot v]\nn\\
& &\cdot\frac{-i}{(p^2+m^2-i\ve)T^2}\exp\{ -i\int d\t
[\l C_{21}(\t)+M_2\frac{p}{m}\cdot u]\}. \nn\\
\label{zxc}
\eE
Since $C_{21}$ has no field dependency it can be extracted out of the
functional integral and with the
integration over $M_2$ (\ref{zxc}) becomes
\be
e^{-i\l\int d\t C_{21}}\int dp\tilde D u\tilde D v\d[u\cdot v]
\frac{-i}{(p^2+m^2-i\ve)T^2}\d[\int d\t \frac{p}{m}\cdot u].
\label{mmm}
\ee

Note that the integrand does not have $v$ field dependency thus integral
over $v$ part also can be decoupled to be absorbed into the normalization
factor. Moreover if $u$ is periodic, it can be mode-expanded thus only the
zero mode $u_0$ remains in the delta functional
$\d[\int d\t \frac{1}{m}p\cdot u]=\d(\frac{T}{m}p\cdot u_0).$ This implies
that along with the integration over $u$ it gives the order $T^{-1}$ which
together with $T^{-2}$ in (\ref{mmm}) can be exactly cancelled by the $T^3$
order of the ghost part (\ref{gauss}).

%\section{Conclusion}
We conclude this letter with some remarks. There have been many studies
on the BRST quantization of the relativistic
spinning particle. In \cite{gat} the massive case is treated, in \cite{pol}
is dealt with the massless case and in \cite{pie} the extended
supersymmetric case is considered.
However in most cases, they use the pseudo-classical
formulation.

In this letter, we have worked the same for
the `classical' system with the relaxed constraint
$\frac{1}{m}p_\m\L^\m{}_\n +t_\n=0$. In the final result (\ref{mmm}), we note
that the BRST invariance
assures the cancellation of the unphysical $T$ dependency of the propagator.
Further the spin factor is the same as in \cite{pol1}. The delta
functional $\d[\int d\t \frac{1}{m}p\cdot u]$ says that only those $u$
satisfying $\int d\t \frac{1}{m}p\cdot u=0$ contribute to the spin factor and
becomes some volume factor upon integration. In such a case
$\frac{1}{m}p\cdot u$ may not vanish for all time, which is just compatible
with the relaxed constraint $\frac{1}{m}p_\m\L^\m{}_i=-t_i.$ The same
is true for $v$ since we can change the role of $u$ and $v$ by the
integration by parts in (\ref{euv}). We can say, therefore, the model
proposed in \cite{cho} describes a spinning particle in the `classical'
fashion comparable with the `pseudo-classical' formulation.

\end{document}